\documentstyle[12pt,epsfig,openbib]{article}
\begin{document}

\begin{titlepage}

\begin{center}
{\Large \bf Two-Proton Correlations near Midrapidity in $p+Pb$ and $S+Pb$
Collisions at the CERN SPS}
\end{center}
\bigskip

{\small
H.~B\o ggild$^{1}$, J.~Boissevain$^{2}$, L.~Conin$^{3}$,
J.~Dodd$^{4}$, B.~Erazmus$^{3}$, S.~Esumi$^{5,a}$,
C.W.~Fabjan$^{6}$, D.E.~Fields$^{2,b}$, A.~Franz$^{6,c}$,
K.H.~Hansen$^{1,\dag}$, E.B.~Holzer$^{6}$, T.J.~Humanic$^{7}$,
B.V.~Jacak$^{8}$, R.~Jayanti$^{7,9}$,
H.~Kalechofsky$^{7,9}$, Y.Y.~Lee$^{7,9}$,
M.~Leltchouk$^{4}$, B.~L{\"o}rstad$^{10}$,
N.~Maeda$^{5,d}$, L.~Martin$^{3}$, A.~Medvedev$^{4}$, 
A.~Miyabayashi$^{10}$, M.~Murray$^{11}$,
S.~Nishimura$^{5,e}$, G.~Paic$^{3,f}$, 
S.U.~Pandey$^{7,9,g}$, F.~Piuz$^{6}$, J.~Pluta$^{3,h}$,
V.~Polychronakos$^{12}$, M.~Potekhin$^{4}$,
G.~Poulard$^{6}$, A.~Sakaguchi$^{5,i}$,
M.~Sarabura$^{2}$, J.~Schmidt-S\o rensen$^{10}$,
M.~Spegel$^{6}$, J.~Simon-Gillo$^{2}$,
W.~Sondheim$^{2}$, T.~Sugitate$^{5}$, 
J.P.~Sullivan$^{2}$, Y.~Sumi$^{5}$,
H.~van~Hecke$^{2}$, W.J.~Willis$^{4}$, 
K.~Wolf$^{11,\dag}$ and N.~Xu$^{2,j}$\\}

\begin{center}
(\small The NA44 Collaboration)
\end{center}
 
\vspace*{2.0cm}
\begin{flushleft}
{\footnotesize
$^{1}$ Niels Bohr Institute, DK-2100 Copenhagen, Denmark.\\
$^{2}$ Los Alamos National Laboratory, Los Alamos, NM 87545, USA.\\
$^{3}$ SUBATECH, Nantes 44070, France.\\
$^{4}$ Columbia University, New York, NY 10027, USA.\\
$^{5}$ Hiroshima University, Higashi-Hiroshima 724, Japan.\\
$^{6}$ CERN, CH-1211 Geneva 23, Switzerland.\\
$^{7}$ Ohio State University, Columbus, OH 43210, USA.\\
$^{8}$ SUNY Stony Brook, Stony Brook, NY 11794, USA.\\
$^{9}$ University of Pittsburgh, Pittsburgh, PA 15260, USA.\\
$^{10}$ University of Lund, S-22362 Lund, Sweden.\\
$^{11}$ Texas A\&M University, College Station, TX 77843, USA.\\
$^{12}$ Brookhaven National Laboratory, Upton, NY 11973, USA.\\
$^{a}$ Now at Heidelberg University, Heidelberg, D-69120, Germany.\\
$^{b}$ Now at University of New Mexico, Albuquerque, NM 87185, USA.\\
$^{c}$ Now at Brookhaven National Laboratory, Upton, NY 11973, USA.\\
$^{d}$ Now at Florida State University, Tallahassee, FL 32306, USA.\\
$^{e}$ Now at University of Tsukuba, Ibaraki 305, Japan.\\
$^{f}$ Now at Ohio State University, Columbus, OH 43210, USA.\\
$^{g}$ Now at Wayne State University, Detroit, MI 48201, USA.\\
$^{h}$ Permanent address: Warsaw University of Technology, Koszykowa 75, 
00-662 Warsaw, Poland.\\
$^{i}$ Now at Osaka University, Osaka 560, Japan.\\
$^{j}$ Now at Lawrence Berkeley National Laboratory, Berkeley, CA 94720, 
USA.\\
$^{\dag}$ Deceased.\\}
\end{flushleft}

\newpage

\begin{abstract}
Correlations of two protons emitted near midrapidity in $p+Pb$ collisions at
450 GeV/$c$ and $S+Pb$ collisions at 200$A$ GeV/$c$ are presented, as measured
by the NA44 Experiment. The correlation effect, which arises as a result of 
final state interactions and Fermi-Dirac statistics, is related to the 
space-time characteristics of proton emission. 
The measured source sizes are smaller than the size of the target lead nucleus 
but larger than the sizes of the projectiles. A dependence on the 
collision centrality is observed; the source size increases with decreasing 
impact parameter. 
Proton source sizes near midrapidity appear to be smaller than those of pions 
in the same interactions. 
Quantitative agreement with the results of RQMD (v1.08) simulations is found 
for $p+Pb$ collisions. For $S+Pb$ collisions the measured correlation effect 
is somewhat weaker than that predicted by the model simulations, implying 
either a larger source size or larger contribution of protons from long-lived
particle decays.
\end{abstract}
 
\begin{flushleft}
\end{flushleft}

\end{titlepage}
 
\section{Introduction}
 
Correlations of identical particles emitted with 
similar velocities are commonly used as a tool to measure 
the space-time development of the emission process
in particle and heavy-ion collisions.
The technique is most often used for bosons, for which the effect
of quantum (Bose-Einstein) statistics is the only source of correlations
(photons), or plays a dominant role (pions and kaons). Other effects, 
in particular those due to the Coulomb interaction for pairs of charged 
particles, are usually treated as a correction.
In the case of protons, final state interactions, due to strong and 
Coulomb forces, typically dominate the effect of quantum statistics and
determine the form of the correlation function. 
A negative correlation effect, due to Fermi-Dirac statistics and Coulomb
repulsion, competes with the positive (attractive) correlation due to
the strong interaction, giving rise to a characteristic ``dip-peak'' structure
(minimum at zero, maximum at $\approx$ 20 MeV/$c$ 
proton momentum in the pair rest frame)
in the two-proton correlation function. The height
of the correlation peak decreases as the proton source size increases.
All effects are strongly sensitive to the space-time parameters 
of the emission process \cite{Koonin,Lednicky,Pratt,Ghisal}.

Correlations of protons emitted with small relative momenta were observed for 
the first time in $\pi^{-}+Xe$ interactions at 9 GeV/$c$ \cite{Siemiarczuk}.
Currently, two-proton correlations are used extensively in low energy 
(of order 10-1000 MeV/$c$) 
heavy-ion physics 
\cite{Pochodzalla,Chen,Gustafsson,Gong}
where the sizes measured for small proton momenta 
are comparable to or exceed the size of the larger (usually target) nucleus.
A similar result was obtained at much higher CERN SPS energies 
in the target fragmentation region, where a two-proton correlation analysis
yielded source sizes compatible with the sizes of the target nuclei 
\cite{Awes}.
A decrease of the measured size with increasing 
proton momenta is reported in many papers however, and for very
different projectile energies \cite{Pochodzalla,Chen,Gong,Bartke1,Azimov}.
Several authors have also reported a dependence on collision impact parameter
in low energy heavy-ion interactions \cite{Gustafsson,Lisa}.
The observed trends in the measured sizes may be related to differences 
in the proton emission time \cite{Chen,Gong}, 
and may also reflect the reaction dynamics,
which can produce correlations between the momentum and position of the 
emitted particles \cite{Fields}.

In spite of extensive analysis of pion and kaon correlations in  
relativistic heavy-ion collisions, data for two-proton correlations are 
scarce. 
This paper reports the first measurement of proton-proton correlations at 
midrapidity at CERN SPS energies.
The origin of protons emitted in relativistic heavy-ion 
collisions differs from that of lighter particles. Protons are
constituent parts of the projectile and target nuclei, unlike pions 
and kaons which are produced in the collision. The different dynamics of
proton emission may therefore be reflected in different source sizes.
At the SPS, the expected net baryon-free region at midrapidity for the 
highest collision energies is not yet achieved. There is still 
significant stopping, with the associated rapidity shifts of projectile
and target nucleons leading to an increase in the proton density at
midrapidity \cite{Bearden1,Bearden2}.
Consequently, the ratio of antiprotons to protons at midrapidity
is approximately 0.1 \cite{Bearden1}. 
The relatively high degree of stopping implies that secondary interactions
of pions, kaons and other produced particles with nucleons are important, 
and may influence the size of the proton emission region.
The relation between the proton source size, the dimensions of the colliding
nuclei, and the corresponding source sizes for pions and kaons at midrapidity
for CERN SPS energies is still an open question.
In this context, the results of the two-proton correlation analysis 
described here provide complementary information
to existing data from two-boson correlations which can be useful in 
understanding the underlying dynamics of the collisions
\cite{Boggild2,Beker1,Beker2,Boggild3,Alber1,Alber2}.

The paper presents the results of a proton-proton 
correlation analysis for collisions of two different 
projectiles, protons and sulphur nuclei, with a lead target. 
In the first case, the projectile consists of a single proton, 
and the measured pairs contain mainly protons from the target 
or protons produced in the interaction. In the second case, the 
correlated pairs contain both projectile and target participants 
as well as produced protons. 
The relative role of these different sources will depend on the
impact parameter of the collisions, i.e. the centrality selection.
This dependence is examined by an analysis of the correlation effects as a
function of charged particle multiplicity in the collision.
Finally, the experimental correlation functions are compared to predictions
obtained from both a Gaussian source model and from the
RQMD event generator, version 1.08 \cite{Sorge}.

\section{Experiment}

The NA44 detector is described in \cite{Boggild2}. The spectrometer is 
optimized to measure single- and two-particle distributions of charged 
hadrons produced near midrapidity in $pA$ and $AA$ collisions. 
The momentum acceptance is $\pm$ 20\% of the nominal momentum setting, and
only particles of a fixed charge sign are detected for a given spectrometer
setting. 

The data used for this analysis were taken at the 6 GeV/$c$ setting
with the spectrometer axis at 44 mrad with respect to the incident beam.
The corresponding acceptance for protons is 2.35 $< y <$ 2.70
and 0.0 $< p_{T} <$ 0.66 GeV/$c$ ($\langle p_{T} \rangle \approx$ 230 MeV/$c$).
Scintillator hodoscopes are used for tracking, and also provide time-of-flight
information with $\sigma_{\rm TOF} <$ 100 ps. The momentum resolution
$\delta p/p$ is approximately 0.2\%. Two threshold Cerenkov detectors in 
conjunction with the time-of-flight (mass-squared) information provide particle
identification. 
Particle contamination is less than 1\% for the data
shown here. 
An interaction trigger is provided by two rectangular 
scintillator paddles sitting downstream of the target and covering 
approximately 1.3 $< \eta <$ 3.5. Offline, a silicon pad multiplicity 
detector with 2$\pi$ azimuthal coverage over the range 1.5 $< \eta <$ 3.3 
is used for event characterization.

Target thicknesses of 10mm and 5mm were used for the $p+Pb$ and $S+Pb$ data,
corresponding to approximately 5.9\% and 5.0\% interaction lengths 
respectively. Multiple scattering in the targets, combined with the 
spectrometer momentum resolution, lead to a resolution in $k^{*}$, the particle
momentum in the rest frame of the pair, of $\sigma(\delta k^{*}) \approx$ 9
MeV/$c$. The experimental trigger required an interaction in the target, 
at least two tracks in the spectrometer, and a veto on pions (and lighter
particles) in the spectrometer acceptance. No centrality selection was
made in the trigger, however the two-track requirement itself biases the
data towards high-multiplicity events.
Offline track reconstruction was followed by track quality cuts, including
rejection of close-by tracks, and particle identification cuts to select
proton pairs. Additional cuts to reject multiple interactions in the target
(pileup) and require clean events were made.
The final event samples are approximately 8k pairs for the
$p+Pb$ data and 15k pairs for the $S+Pb$ interactions.

\section{Correlation Analysis}

The experimental correlation function is constructed from identified
two-proton events. Due to limited statistics, only a one-dimensional
analysis has been performed.
The correlation function is calculated as:
 
\begin{equation}
C(k^{*}) = \frac{N_{corr}(k^{*})}{N_{uncorr}(k^{*})}
\end{equation}

where $k^{*}$ is the particle momentum in the rest frame of
the pair: $k^{*} = \frac{\sqrt{-Q^2}}{2}$, 
$Q \equiv \{q_{0}, \vec{q}\} = p_{1}-p_{2}$ and  $p_{1}, p_{2}$ are
the particle 4-momenta. 
The correlated, $N_{corr}$, and uncorrelated, $N_{uncorr}$,
proton pairs are taken from the same and different events
respectively. 
The stability of the resulting correlation function under the applied cuts 
is studied by varying each cut (such as the mass-squared selection, and the
hodoscope multiplicity) individually. In all cases, the correlation function
data points are within one standard deviation of their nominal values. The
stability with respect to different sampling procedures for producing the 
mixed-event distribution is also tested, and introduces no systematic
deviations.

The experimental correlation functions presented here do not 
contain any correction factors. Such corrections are often made in
boson correlation measurements to take account of the effects of finite 
detector acceptance and momentum resolution, residual correlations 
in the background distribution \cite{Zajc}, and the Coulomb interaction,
thereby recovering the ``ideal'' correlation function in which correlations
are due purely to quantum statistics.
Rather than correct the experimental distributions, theoretical
correlation functions based on model predictions have been generated, and
modified to include each of these effects. The results of these simulations
are then compared directly to the measured correlation functions.

The theoretical correlation function is generated by selecting proton
pairs from a given source model, calculating the associated 
weight due to quantum statistics and final state interactions (strong and
Coulomb) \cite{Lednicky,Pratt}, 
and propagating the particles through a realistic simulation of
the experimental detector, including all instrumental and acceptance effects.
A full calculation of Coulomb effects is used rather than the Gamow factor
approximation.
Data cuts
are then made in
exactly the same way as for the experimental data.
For this analysis, two models have been used as inputs for the simulation:
a Gaussian source model and the RQMD event generator.

In the case of the Gaussian model, a Gaussian density distribution
$\rho(\vec{r},t) = \frac{1}{(2\pi)^{2}r_{o}^{3}\tau} 
e^{-\frac{\vec{r}^{2}}{2r_{o}^{2}} - \frac{t^{2}}{2\tau^{2}}}$
is assumed 
for both the spatial and temporal extent of the source \cite{Lednicky}. 
It is assumed that protons are emitted 
by sources moving in the longitudinal direction with the velocities of 
the proton pairs (the Longitudinally Co-moving Source, LCMS). In this way the 
fast longitudinal motion of proton sources at mid-rapidity is taken into 
account.
Because the rapidity acceptance in this measurement is narrow 
(2.35 $< y <$ 2.70),
the results assuming the LCMS should not differ significantly from the results
assuming a fixed frame of reference centered at about $y$ = 2.5.
Since a one-dimensional analysis is performed 
here, a spherically-symmetric source parametrized by equal $r_{o}$ radius 
and $\tau$ time parameters is assumed. The RMS radius in this 
parametrization is given by 
$\sqrt{3}r_{o}$. 

In order to compare with RQMD predictions, the theoretical correlation 
function for pairs of 
protons from generated events is calculated \cite{Pratt}, 
and the appropriate weighting
for different impact parameter collisions is taken into account by matching
the multiplicity distributions of the data and model.
This matching shows that, in contrast to the $p+Pb$ data where
only the highest multiplicity events are selected by the trigger, 
the $S+Pb$ data are only weakly biased toward central collisions.

In calculating correlation functions for comparison to data, three 
factors are particularly important: the admixture of indirect
protons coming from hyperon 
(mainly $\Lambda \rightarrow \pi^{-}p$) decays, the acceptance 
and resolution of the experiment, and the residual correlations 
arising in the single particle background distributions. 

Weak decays of strange baryons are a significant source of protons
and contribute to the yields measured in the NA44 spectrometer.
The influence of the admixture of indirect protons on the
shape of correlation function has been studied using data from the 
RQMD and Venus (v5.21) \cite{Werner} event generators, combined with a
detailed simulation (GEANT) of the detector.
The two models give similar results: about 22\% of protons measured
in the spectrometer come from the decay of $\Lambda$'s in both 
$p+Pb$ and $S+Pb$ collisions, and cannot be distinguished from
direct protons.
This contribution has only a weak $p_{T}$ dependence, and is greater
at low $p_{T}$.
In order to take into account this non-correlated contribution to
coincident pairs, a fraction of 22\% of ``decay'' protons is included
when calculating the correlation function, giving about 39\% of
uncorrelated pairs in the two-particle sample. These pairs are assigned 
a weight of unity in the correlation function calculation.
This contribution significantly reduces the magnitude
of the observed correlation effect, but does not
change the general shape of the correlation function.

The acceptance and resolution of the NA44 spectrometer for protons is 
estimated using a detailed simulation of the detector, including 
systematic effects introduced by the track reconstruction procedure for
close-by pairs of particles.
The $k^{*}$ acceptance for proton pairs is convoluted with the 
detector resolution and the rather complicated shape of the two-proton 
correlation function, and gives rise to an asymmetric distortion of 
the correlation peak. This ``smearing'' is particularly important for
small $k^{*}$, where the expected depletion due to Coulomb and
statistical effects is masked by the experimental resolution.

Residual correlations arise in the mixing procedure because 
the single particle distributions forming the reference (background)
sample are in fact projections of the measured two-particle distribution, 
which includes the correlations.
This effect is taken into account 
by calculating and applying weights which correspond to the mean 
value of the correlation function for all combinations of a given 
particle with all others \cite{Zajc}.  
This residual correlation modifies slightly 
the height of the correlation peak, and induces a slope 
for large values of $k^{*}$.

All three effects are included in the simulation procedure. 
In Fig. 1 they are demonstrated separately in order to illustrate 
their particular features and the relative significance 
of each of them.

\section{Results}

The experimental proton-proton correlation functions for $p+Pb$ and 
$S+Pb$ collisions are presented in Fig. 2. 
The correlation functions have been normalized such that $C(k^{*}) = 1$ for
$80 < k^{*} < 160$ MeV/$c$.
The peak observed in the region of small $k^{*}$ values ($\approx$ 20 MeV/$c$)
can be attributed to the combined effect of the final state 
interaction and Fermi-Dirac statistics. It reflects the space-time 
properties of the proton emission process. 
The two correlation functions have qualitatively the same shape, but the
peak is much more pronounced for the $p+Pb$ data than for the $S+Pb$,
indicating that protons are emitted from a smaller source in the case 
of $p+Pb$ collisions. 
Also shown are the correlation functions and associated source sizes from 
the Gaussian model obtained with a minimum-$\chi^{2}$ fit over the range
$0 < k^{*} < 160$ MeV/$c$.
The fits yield
$r_{o} = \tau$ = 1.42$^{+0.04}_{-0.05}$ fm ($\chi^{2}/N_{DF}$ = 10.3/10)
for $p+Pb$ and 
$r_{o} = \tau$ = 2.65$^{+0.09}_{-0.09}$ fm ($\chi^{2}/N_{DF}$ = 9.6/10)
for the $S+Pb$ data. 
(Note that the $r_{o}$ values have to be multiplied by $\sqrt{3}$ 
to obtain the corresponding RMS radii.)
Fits assuming $\tau$ = 0 give $r_{o}$ parameters which are equal within
errors to those obtained with $\tau = r_{o}$.
This is a consequence of the weak time dependence of the final state 
interaction effects for small particle velocities \cite{Lednicky}; for these
data the mean velocity of the proton pairs in the LCMS frame is $\approx$
0.24$c$.

These results have been obtained assuming that the fraction of measured 
protons from $\Lambda$ decay is that given by the simulations described 
above (22\%). 
While the RQMD and Venus models give consistent results for the 
fraction of protons from hyperon decays in the NA44 acceptance, there 
is a factor of order two discrepancy between the available experimental
data on $\Lambda$ production yields in $S+A$ collisions at CERN
SPS energies \cite{Gazdzicki,Judd}.
The RQMD prediction is roughly midway between the two measurements.
Using the experimental data as an estimate of the uncertainty on the
$\Lambda$ yield (approx. $\pm$ 40\%), and hence on the fraction of 
protons from $\Lambda$ decays, gives
$r_{o} = \tau$ = 1.42$^{+0.16}_{-0.17}$ fm for $p+Pb$ and 
$r_{o} = \tau$ = 2.65$^{+0.19}_{-0.17}$ fm for $S+Pb$. 
The RMS radii of the proton and the sulphur and lead nuclei are 0.88 fm,
3.26 fm and 5.50 fm 
respectively \cite{Brown}. In the frame of this geometrical interpretation,
the measured source sizes are greater than the size of the projectile but 
smaller than that of the target.

In order to analyze the centrality dependence of the correlation 
effect, both the $p+Pb$ and $S+Pb$ data have been divided into three
multiplicity bins, based on the charged-particle multiplicity measured
by the silicon detector. 
The binning is chosen to give roughly equal proton-pair statistics
in each sub-sample.
The correlation functions corresponding to the three multiplicity 
selections are shown in Fig. 3 for $p+Pb$ and $S+Pb$.
A systematic decrease of the correlation effect with increasing multiplicity 
is observed, corresponding to an increase in the size of the emitting volume. 
The $r_{o}$ values indicate the scale of these changes. 

The comparison of the data with the predictions of RQMD is shown 
in Fig. 4. In the case of $p+Pb$ collisions, the experimental data 
are well reproduced by the model. For $S+Pb$, the model gives a 
somewhat larger correlation effect than is observed in the data.
Fitting the $S+Pb$ RQMD prediction with the same Gaussian model (and same
fraction of protons from $\Lambda$ decay) as used for the data yields
$r_{o} = \tau$ = 2.27$^{+0.02}_{-0.02}$ fm, compared to 
$r_{o} = \tau$ = 2.65$^{+0.19}_{-0.17}$ for the data.

\section{Discussion}

A simple, one-dimensional description with Gaussian co-moving sources 
is used to quantify the correlation effects. 
In introducing the co-moving source, it is assumed that particles 
with similar longitudinal velocities are emitted from nearby space-time 
points, such as occurs when fast longitudinal expansion takes place.
In this case, the correlation function effectively measures only a 
part of the space-time extent of the emission volume.
For the $S+Pb$ system,
the multiplicity cuts select different impact parameter collisions, and
hence different sizes of the overlapping region between projectile 
and target nucleons. This is reflected in the observed centrality dependence 
of the correlation functions.
Comparison of the measured source sizes with the RMS radii of the projectile 
and the target is also in qualitative agreement with such an interpretation.
Note that in the target fragmentation region the sizes measured by 
two-proton correlations reflect rather the size of the target nucleus 
\cite{Awes}.
In the case of $p+Pb$ collisions, the number of 
projectile participants is always equal to one
and the measured size can be related to the extent 
of the hot region along the path of the proton in the lead nucleus.
The source sizes in this case are much smaller than in the case of the
sulphur projectile. 

These data can be compared to correlation results obtained for 
other particle types near midrapidity and in similar $p_T$ intervals,
which assume the same type of emitting source.
For both systems studied ($p+Pb, S+Pb$), the proton source 
sizes appear to be smaller than those of pions,
especially in the case of the sulphur projectile: 
for $p+Pb$, 
$r_{Ts}(\pi^{+})$ = 2.00 $\pm$ 0.25 fm 
while for $S+Pb$, 
$r_{Ts}(\pi^{+})$ = 4.15 $\pm$ 0.27 fm 
($\langle p_{T} \rangle \approx$ 150 MeV/$c$ in both cases) 
\cite{Beker2,Boggild3}.
The $r_{o}$ values for protons are similar to 
the corresponding kaon source sizes:
for $p+Pb$, 
$r_{Ts}(K^{+})$ = 1.22 $\pm$ 0.76 fm 
while for $S+Pb$, 
$r_{Ts}(K^{+})$ = 2.55 $\pm$ 0.20 fm 
($\langle p_{T} \rangle \approx$ 240 MeV/$c$ in both cases) 
\cite{Beker1}.
(Note that for the
three-dimensional boson source parameters, multiplication by a
factor of $\sqrt{3}$ is also necessary to obtain RMS sizes). 
The Gaussian source model used in these analyses describes sources for 
which the momenta and positions of the emitted particles are 
independent in the LCMS.
Collision dynamics may, however, induce correlations between the
measured source size and the transverse momentum of the particles.
In particular, a decrease of the measured size with increasing 
transverse mass $m_{T}$ (where $m_{T} = \sqrt{p_{T}^{2} + m^{2}}$ ),
observed in common for pions and kaons,
may indicate the collective expansion of an equilibrated system
formed during the collision \cite{Fields,Beker2}. 
The smaller source sizes for the two-proton data ($\langle m_{T} \rangle 
\approx$ 970 MeV) compared to the two-boson data \cite{Beker1,Beker2,Boggild3} 
are in qualitative agreement with such an $m_{T}$ dependence.

The contribution of indirect protons coming from hyperon 
decays is also a parameter in the description of the 
experimental results, and can influence the values of the extracted 
source sizes. 
The experimental uncertainty on the $\Lambda$ yield is described above,
and results in a systematic error of approximately 0.2 fm on the extracted 
$r_{o}$ parameters. Of the other hyperons, Monte Carlo studies show that
only the $\Sigma^{+}$ also contributes to the proton yield measured 
in the NA44 spectrometer. RQMD and Venus predict that the number of detected 
protons from $\Sigma^{+}$ decay is at most 30\% and 10\%, respectively, 
of that from $\Lambda$ decay - less than the experimental uncertainty on the 
$\Lambda$ yield itself.
Possible uncertainty in the contribution of indirect protons cannot 
significantly alter the conclusion on the size of the proton source
compared to the pion source however, nor on the centrality dependence 
of the observed sizes. 
In order to produce the change in the measured correlation functions between 
the lowest and highest multiplicity bins assuming that the difference is due 
entirely to indirect protons implies that more than 50\% of detected protons 
come from hyperon decays.
However, an enhanced emission of hyperons with respect to that predicted 
by the RQMD model in the case of $S+Pb$ collisions may be responsible for the 
difference between the model prediction and the results of measurements.

\section{Conclusion}

These results demonstrate the first attempt to include baryon-baryon 
correlations in the analysis of the space-time dynamics of particle 
emission at midrapidity at CERN SPS energies. 
The measured source sizes increase with the mass of the projectile 
and with the collision multiplicity. 
Within the context of a Gaussian source model, the proton radius
parameters are smaller than the size of the target nucleus but larger
than the sizes of the projectiles.
Comparing to other particle types, the proton size parameters are smaller
than those for pions, and similar to those for kaons, measured in the 
same collisions.
(Note that collision dynamics may induce a dependence of the measured 
source sizes on the momentum of the emitted particles.)
Good agreement with the results of RQMD (v1.08) simulations is seen for
$p+Pb$ collisions. For $S+Pb$ collisions, the measured correlation
effect is somewhat weaker than that predicted by the model simulations.  

\section{Acknowledgements}

The NA44 Collaboration wishes to thank the staff of the CERN PS-SPS 
accelerator complex for their excellent work. We thank the technical
staff at CERN and the collaborating institutes for their valuable
contributions. We are also grateful for the support given by the
Science Research Council of Denmark; the Japanese Society for the
Promotion of Science, and the Ministry of Education, Science and
Culture, Japan; the Science Research Council of Sweden; the US
Department of Energy and the National Science Foundation. 
We thank Dr. H. Sorge for the use of the RQMD code; Dr. K. Werner
for the Venus code; and Drs. R. Lednicky and S. Pratt for making
their correlation codes available to us, and for fruitful discussions.

\newpage

\begin{figure}
   \begin{center}
      \mbox{\epsfxsize=16cm\epsffile{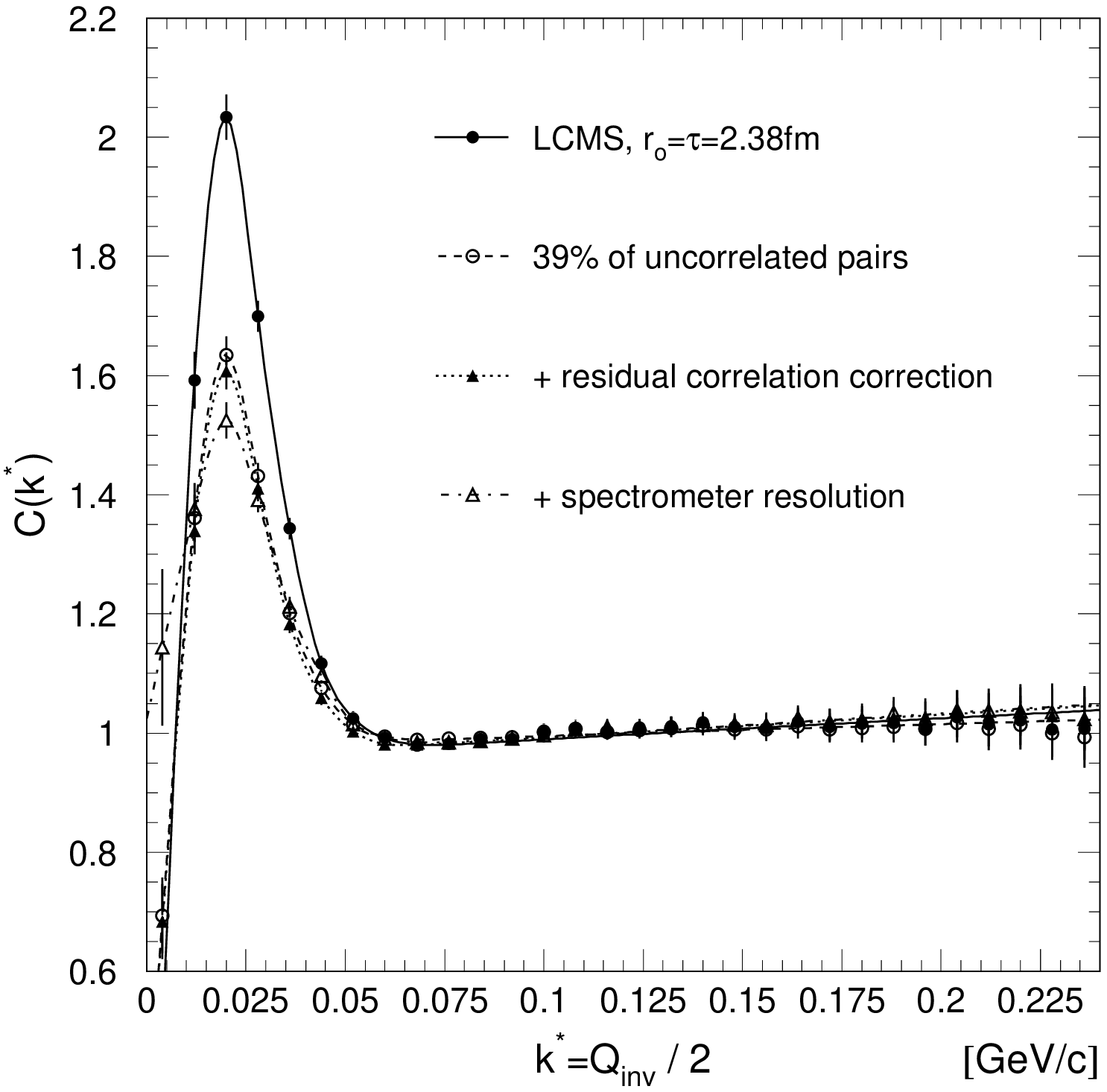}}
      \caption{
       The influence of different experimental factors on the 
       form of the measured correlation function. The example of a
       Gaussian source with $r_{o} = \tau$ = 2.38 fm is shown, where
       the effect of each contribution has been added cumulatively. 
       }
   \end{center}
\end{figure}

\begin{figure}
   \begin{center}
      \mbox{\epsfxsize=16cm\epsffile{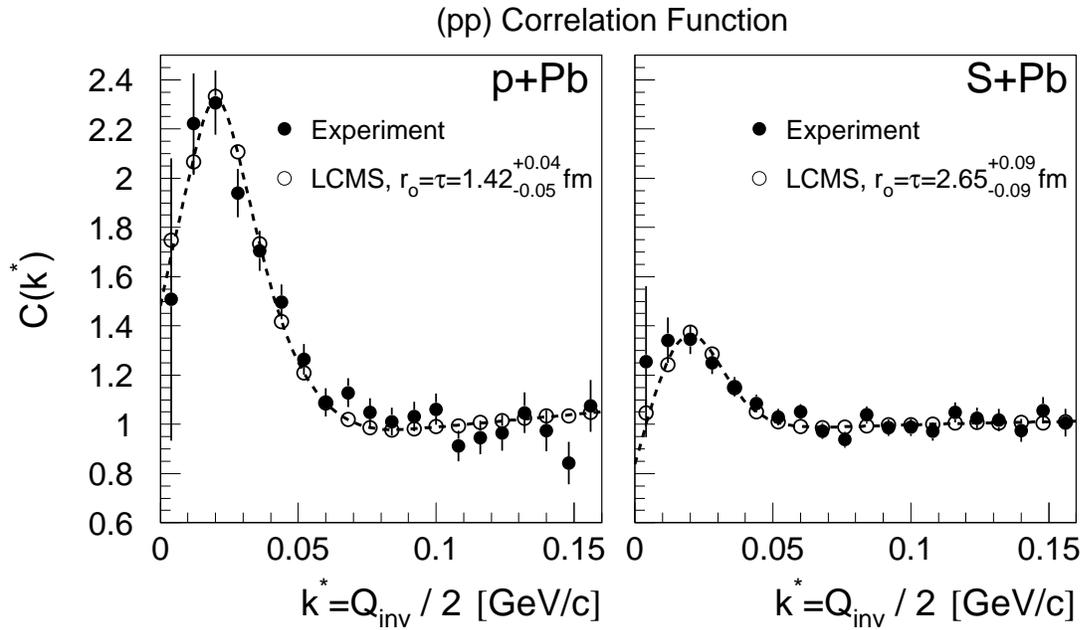}}
      \caption{
       Two-proton correlation function for $p+Pb$ (left) and $S+Pb$ (right) 
       collisions. The open points and dashed lines show the predicted 
       correlation functions for longitudinally co-moving Gaussian sources 
       with the radii indicated in the figure.               
       }
   \end{center}
\end{figure}

\begin{figure}
   \begin{center}
      \mbox{\epsfxsize=16cm\epsffile{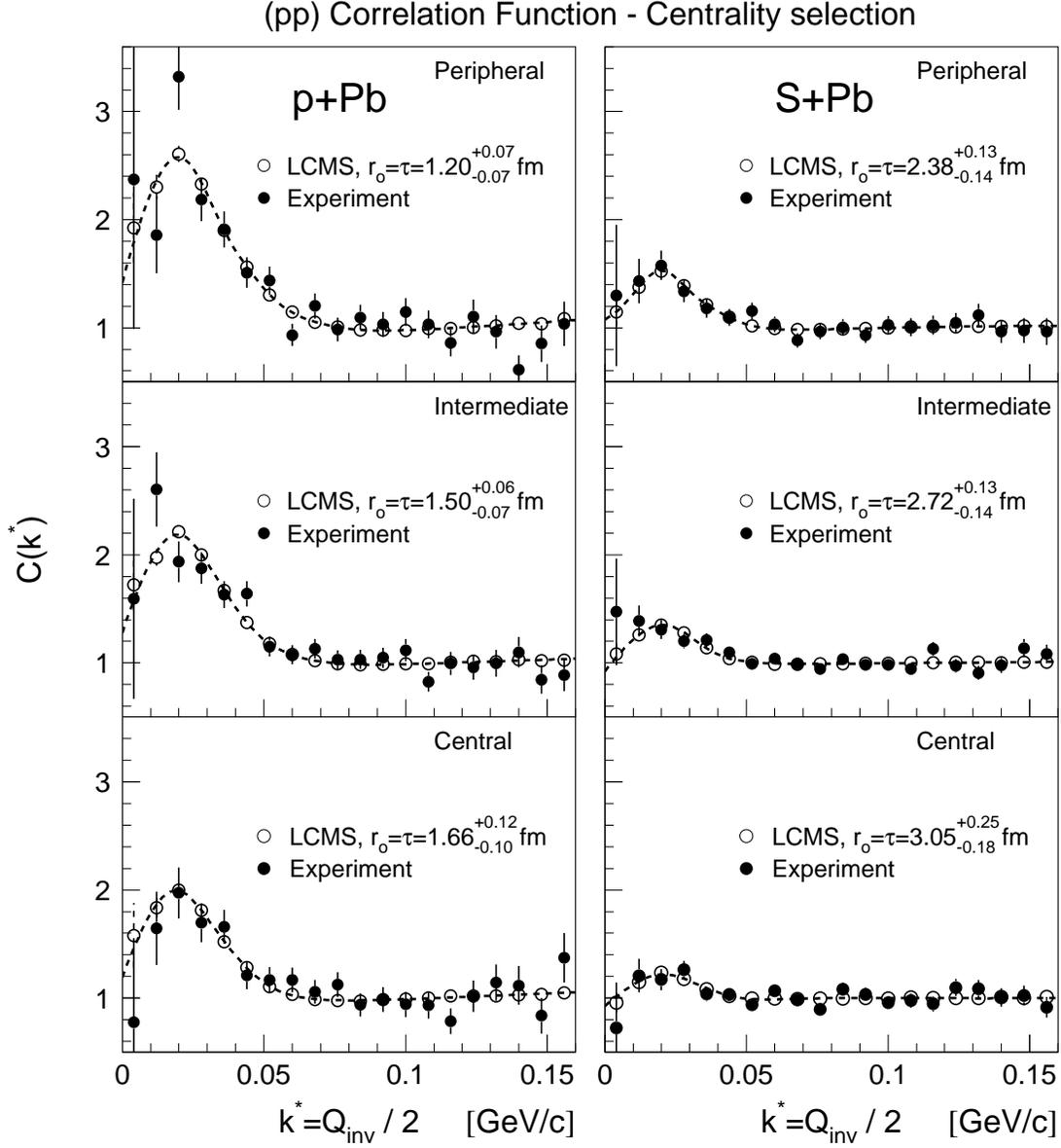}}
      \vspace*{-2.0cm}
      \caption{
       Two-proton correlation function for $p+Pb$ (left) and $S+Pb$ (right) 
       collisions for three different multiplicity selections. The open points
       and dashed lines show the form of correlation function for 
       longitudinally co-moving Gaussian sources with the radii indicated 
       in the figure. 
       }
   \end{center}
\end{figure}

\begin{figure}
   \begin{center}
      \mbox{\epsfxsize=16cm\epsffile{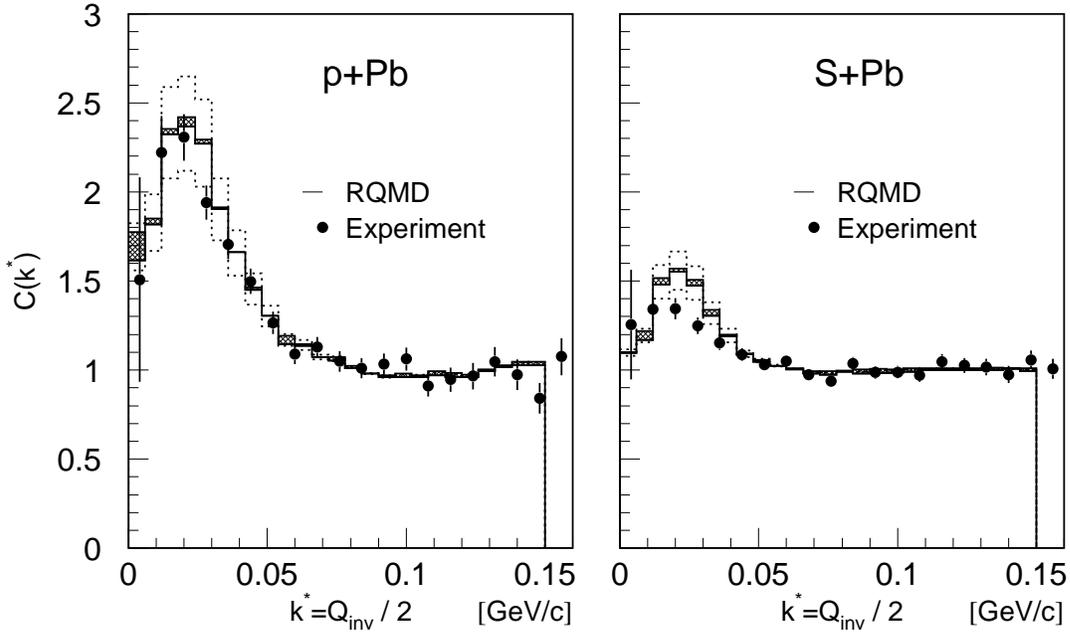}}
      \caption{
       Comparison of the experimental correlation functions with
       predictions of the RQMD (v1.08) model for $p+Pb$ (left) and $S+Pb$ 
       (right) collisions.
       The hatched error band on the RQMD predictions includes the systematic 
       error due to the uncertainty in the spectrometer resolution and the 
       uncertainty on the impact parameter weighting obtained by matching 
       the multiplicity distributions of the data and the model. The 
       non-hatched error band reflects the experimental uncertainty on the
       measured $\Lambda$ yields \cite{Gazdzicki,Judd}. Statistical 
       errors are negligible.
      }
   \end{center}
\end{figure}

\end{document}